\documentclass[11pt]{article}
\usepackage{float}
\usepackage{latexsym}
\usepackage{graphicx}
\usepackage{enumerate} \usepackage{delarray}
\usepackage{tabularx} \usepackage{bm} \usepackage{amssymb}
\usepackage{cite}

\usepackage[small,labelfont=bf,labelsep=space]{caption}
\makeatletter
\def\@cite#1{\textsuperscript{\tiny{#1}}}
\makeatother
\begin{document}

\makeatletter
\renewcommand{\@biblabel}[1]{#1.}
\makeatother

\title{{\LARGE Efficient immunization strategies to prevent financial contagion}} 

\author{Teruyoshi Kobayashi\footnote{Graduate School of Economics, Kobe University, 2-1 Rokkodai-cho, Nada-ku, Kobe 657-8501, Japan. E-mail: kobayashi@econ.kobe-u.ac.jp.} \and  Kohei Hasui\thanks{Graduate School of Economics, Kobe University.}}
 \date{\large {December, 2013}} \maketitle
\begin{abstract}
Many immunization strategies have been proposed to prevent infectious viruses from spreading through a network. In this work, we study efficient immunization strategies to prevent a default contagion that might occur in a financial network. An essential difference from the previous studies on immunization strategy is that we take into account the possibility of serious side effects. \textit{Uniform immunization} refers to a situation in which banks are \lq\lq{}vaccinated\rq\rq{} with a common low-risk asset. The riskiness of immunized banks will decrease significantly, but the level of systemic risk may increase due to the de-diversification effect. To overcome this side effect, we propose another immunization strategy, called \textit{counteractive immunization}, which prevents pairs of banks from failing simultaneously. We find that counteractive immunization can efficiently reduce systemic risk without altering the riskiness of individual banks.

\end{abstract} \thispagestyle{empty} \newpage \pagenumbering{arabic}


Since the global financial crisis of 2007-2009, many researchers in various fields of natural and social sciences, such as physicists, ecologists and economists, have been tackling the question of how to reduce financial systemic risk.\cite{beale11,caccioli12,gai10,gai11} Many of those studies are based on network theory, which has been used extensively to explore complex systems since the early 2000\rq{}s. 

Percolation is one of the most frequently used concepts in thinking about the fragility of networks. Percolation theory provides a way to reveal the threshold of the number of removed nodes above which the giant component disintegrates.\cite{callaway00,schwartz02,boguna05,squires12} Applying this theory, researchers examined the fragility of various types of actual complex networks, such as the Internet, road networks, and power grids.\cite{albert00,albert04,wu06} Percolation theory also gives us some useful insights into the question of how to control the way in which infectious viruses spread through a network. Many studies proposed various immunization strategies to make the percolation threshold as small as possible.\cite{schwartz02,gallos07,schneider11,cohen03,chen08}

 Systemic risk can be viewed as a fear of default contagion. Bank defaults can spread through a financial network in a manner similar to the spread of infectious diseases through a social network. The idea of efficient immunization  would therefore give us a clue to the question of how to reduce the likelihood of default contagion. 
In this work, we explore efficient bank-immunization strategies by using a simple model of the interbank network (see Methods). 

We consider two different types of \lq\lq{}vaccine\rq\rq{}. One is a very-low-risk asset, possibly government bonds or cash. Since only a few of all the types of widely traded assets are categorized as very-low-risk assets, we assume that one such asset is used as a common vaccine. 
The other vaccines are pairs of negatively correlated assets, which could be interpreted as opposite positions in risky assets.  We assume that the riskiness of each of the negatively correlated assets is the same as that of the uncorrelated assets that banks originally held before immunization. 
We call the former type \textit{uniform immunization} and the latter \textit{counteractive immunization} (see Methods).
 Counteractive immunization does not change the riskiness of individual banks, but Kobayashi\cite{kobayashi13} showed that assigning negatively correlated assets to highly \lq\lq{}infective\rq\rq{} banks would  reduce systemic risk.

Even if a common vaccine is used, the effects of immunization will differ if the order of vaccination varies. We show that the likelihood of a large-scale contagion, what is called a \textit{financial crisis}, can be reduced very efficiently if the order of immunization is based on PageRank. At the same time, however, uniform immunization leads to an undesirable situation in which the number of bank defaults during a crisis tends to increase as immunization proceeds. In other words, uniform immunization will make the financial market \textit{robust-yet-fragile}. 
We also examined some different orders of vaccination  that are based on eigenvector centrality, degree centrality, node-betweenness centrality, and the eigenvector-based index proposed by Restrepo et al.\cite{restrepo08} It turns out that the PageRank-based vaccination strategy is most efficient in terms of reducing the likelihood of financial crises.
 It is shown that the undesirable trade-off between the likelihood and the size of a crisis does not arise under counteractive immunization. 

 As an objective measure of systemic risk, we employ the $\chi$-th moment of the number of defaulted banks.\cite{beale11,kobayashi13}
In addition, to see the effects of immunization on the connectivity of the financial network, we also show the likelihood that the giant strongly connected component (GSCC) disappears due to the removal of defaulted banks (see Methods). These two measures are supposed to capture different aspects of the undesirable influence of bank defaults. Nevertheless, we find that the two measures are actually closely related and virtually coincide under a certain value of $\chi$. 

Finally, we examine under what circumstances counteractive immunization becomes more effective than uniform immunization.  
We derive the threshold of the degree of social risk aversion, $\chi$, above which counteractive immunization becomes more advantageous than uniform immunization.
\begin{figure}
\includegraphics[width=13cm,clip]{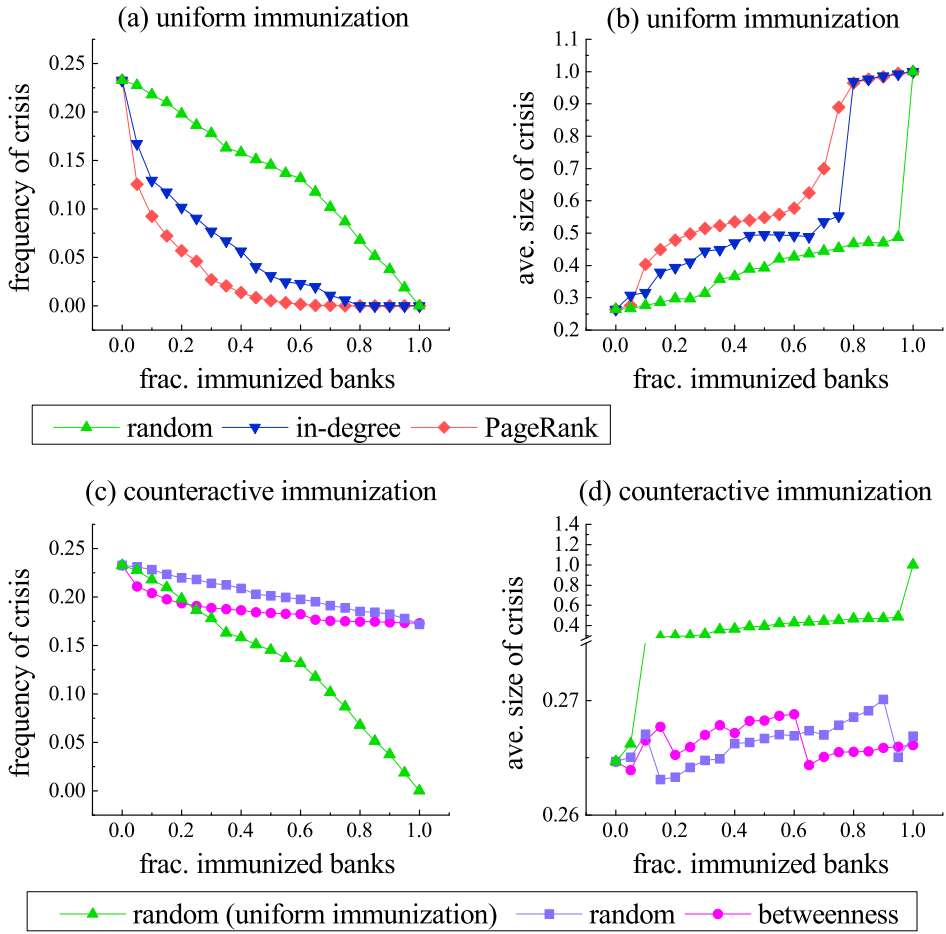}
\caption{{\bf $|$ The frequency and size of financial crises under alternative immunization strategies.}  \lq\lq{}Financial crisis\rq\rq{} is defined as a situation in which at least 5\% of banks go bankrupt. The average of crisis size is taken conditional on crisis event. (a) and (b): The case of uniform immunization.  (c) and (d): The case of counteractive immunization. The outcomes of uniform immunization based on a random vaccination order are also shown for comparison. Returns of external risky assets follow the student t-distribution with degree of freedom 5. The probability of fundamental default, $\delta_r$, is $5/N$ prior to immunization. The probability of fundamental default after uniform immunization, $\delta_s$, is $1/(10N)$. The degree of negative correlation under counteractive immunization, $\rho$, is .6.}
\end{figure}

\section*{Results}
 \subsection*{The likelihood and size of a financial crisis}
   Following Gai and Kapadia\cite{gai10}, a \textit{financial crisis} is defined as a situation in which at least 5\% of banks go bankrupt. Fig.\,1 illustrates the frequency and the conditional average of the size of crises. Some interesting properties are observed.
 First, Fig.\,1a reveals that the uniform immunization strategy based on PageRank is most successful  in terms of reducing the frequency of crises. The least efficient one is random immunization. The reason for this is that a bank with a high PageRank score generally a) has many interbank liabilities (i.e., high in-degree) and b) borrows from lenders that have few interbank assets (i.e., low out-degree). In this model, the fact that a bank borrows a lot means that the bank borrows from many other banks (see Methods). Such a bank is likely to be systemically important because its failure would undermine the balance sheets of many lenders. In addition, the fact that the lenders of bank $i$ have few interbank assets indicates that the lenders\rq{} solvency depends largely on the solvency of bank $i$ since the lenders\rq{} portfolios of interbank assets are not so diversified. 
  
 Secondly, Fig.\,1b states that the average number of banks that fail during a crisis increases as more banks become immunized.  This is because a larger number of banks begin to hold a common low-risk external asset, which reduces the frequency of medium-scale simultaneous defaults (Fig.\,2a).   
 Moreover, Fig.\,1b also shows that a crisis is likely to become more severe under the PageRank-based strategy than under the other immunization strategies. Immunizing banks in the order based on PageRank \lq\lq{}efficiently\rq\rq{} increases the probability that systemically important banks will fail simultaneously, which would lead to a crisis. In other words, the PageRank-based strategy will rapidly shift the financial market toward a \textit{robust-yet-fragile} system.

\begin{figure}
\includegraphics[width=15cm,clip]{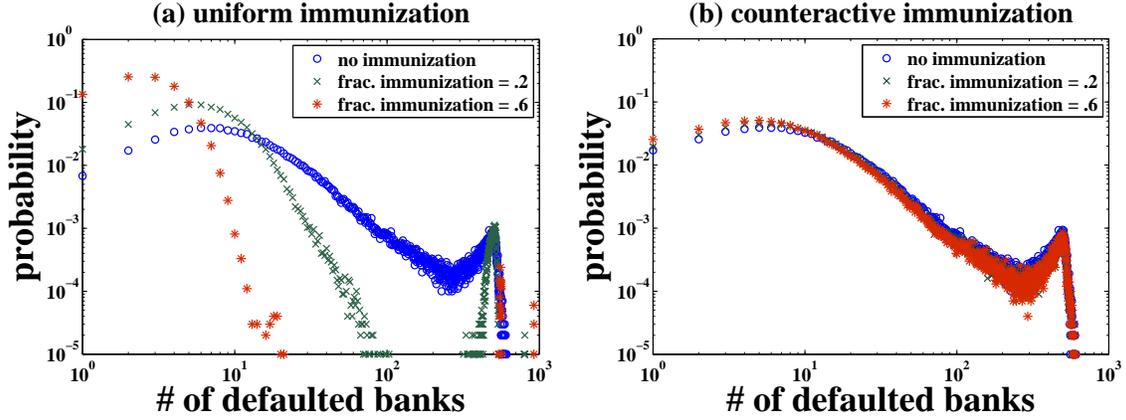}
\caption{{\bf $|$ Histogram of the number of defaulted banks.} (a) Uniform immunization based on PageRank. (b) Counteractive immunization based on edge-betweenness centrality. The figures show that i) uniform immunization decreases the frequency of medium-scale defaults while increasing the likelihood of extremely large-scale defaults. ii) Counteractive immunization slightly reduces the frequency of medium-scale cascades while increasing the likelihood of small-scale defaults. See the caption of Fig.1 for parameter values.}
\end{figure}   

 The corresponding figures for counteractive immunization are shown in Figs.\,1c,d. Counteractive immunization introduces anti-correlations between  two assets that were previously independent of each other without changing the riskiness of individual assets. In contrast to the case of uniform immunization, counteractive immunization decreases the frequency of crises while having an unambiguous influence on the average number of bank defaults at the time of a crisis. 
Assigning negatively correlated assets to pairs of connected banks will avoid medium-scale default cascades by preventing the balance sheets of the immunized banks from moving in sync (Fig.2b). 

 However, the speed at which the frequency of a crisis decreases is much slower under counteractive immunization than under the PageRank-based uniform immunization. This implies a trade-off between frequency and severity.
 The desirability of an immunization strategy thus depends on the extent of social risk aversion, represented by parameter $\chi$. In the following, we clarify how the level of systemic risk under various types of immunization strategies depends on the degree of risk aversion, $\chi$.     

\begin{figure}
\includegraphics[width=15.5cm,clip]{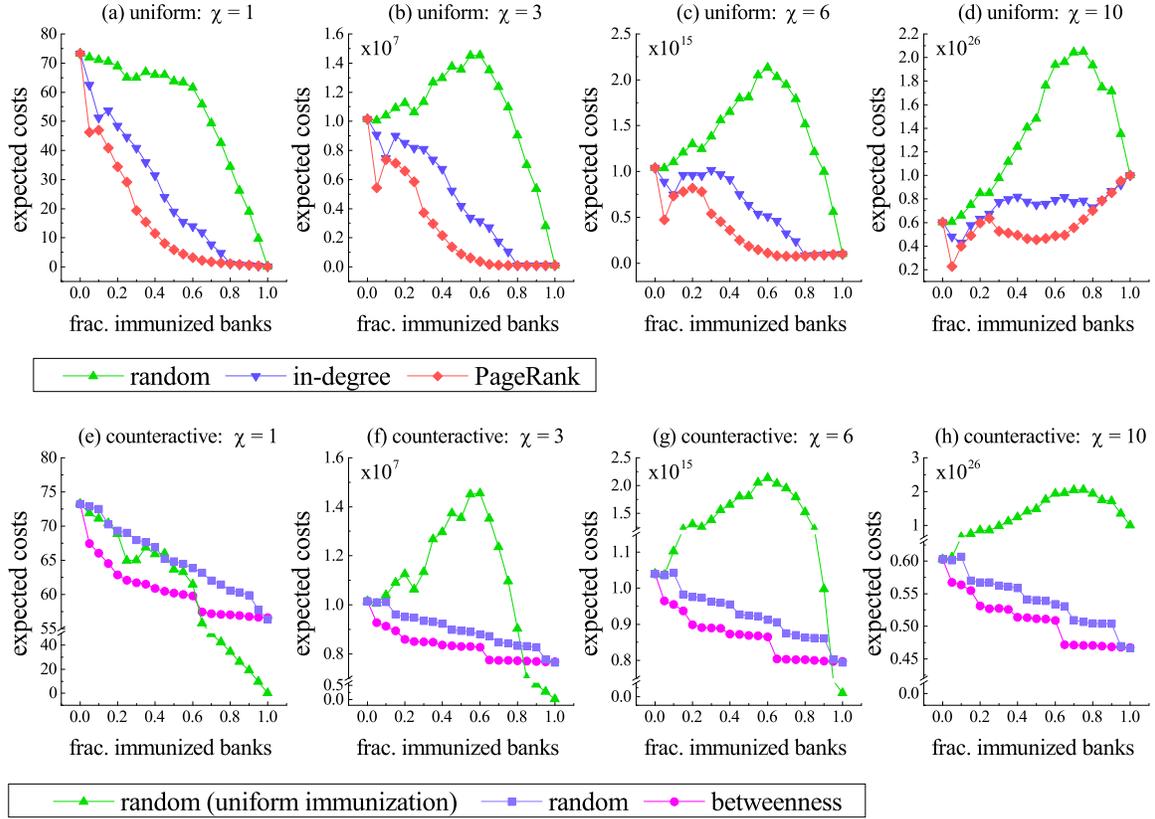}
\caption{{\bf $|$ The relationship between the fraction of immunized banks and the expected costs.} (a)-(d): The case of uniform immunization.  The increases in the expected costs observed in (a) reflect the de-diversification effect. The expected costs tend to become upward-sloping as the value of $\chi$ increases. This reflects the fact that the society is becoming more risk-averse. (e)-(h): The case of counteractive immunization. The outcome of uniform immunization based on a random vaccination order is also shown for comparison. The shape of the curve is little affected by the value of $\chi$ under counteractive immunization. See the caption of Fig.1 for parameter values.}
\end{figure}

\subsection*{Systemic risk}

  Fig.\,3 depicts the relationship between the fraction of immunized banks and the expected costs under four different values of $\chi$. Some important features should be noted.  First, under uniform immunization, the expected costs  may increase even if $\chi =1$, where the expected cost is equal to the average number of defaults. 
 An intuition into why the average number of defaults increases  is that a homogenization of banks\rq{} external assets effectively makes the portfolio of interbank assets less diversified. If every bank has an idiosyncratic external asset, then the holdings of multiple interbank assets contribute to portfolio diversification. However, if borrowers begin to hold the same external assets as their lenders\rq{}, then the effective degree of the lenders\rq{} portfolio diversification will be lowered. Such an effect, what we call the \textit{de-diversification effect}, may increase the average size of default cascades even when the risk of external assets is diminishing.
 On the other hand, asset harmonization itself also has an effect of decreasing contagious defaults. This is because asset harmonization increases the average number of banks that would fail simultaneously due to the loss of external assets only. Thus, if the riskiness of the common asset is sufficiently low, then asset harmonization would reduce the expected costs.  

 Secondly, the expected costs are more likely to become upward-sloping as $\chi$ increases. This is because the degree of convexity of the cost function increases as $\chi$ goes up, reflecting the fact that the society is becoming more risk-averse. For large values of $\chi$, the possibility of a large-scale simultaneous default turns out to be an important source of risk even though its probability of occurrence is quite low. 

 In contrast, Figs.\,3e-h show that counteractive immunization reduces systemic risk virtually monotonically. This is because counteractive immunization can lower the likelihood of medium-scale default cascades without causing increased asset commonality.  A limitation of this strategy is that it does not change the riskiness of a single external asset itself. In fact, assigning anti-correlated assets to a pair of isolated banks can slightly lower the probability that both of the banks will survive. To demonstrate this, suppose that the probability of default is $p$ for each bank and that the correlation coefficient of the asset returns is -1. The probability that both banks will be solvent is then $1-2p$. If asset returns are independent, then the corresponding probability is given by $(1-p)^2$. Since $(1-p)^2 -(1-2p)$ $= p^2$, it follows that counteractive immunization reduces the probability that both banks will survive. This is why counteractive immunization may slightly increase the expected costs in some circumstances. 

\subsection*{Fragility of the financial network}

\begin{figure}
\includegraphics[width=16cm,clip]{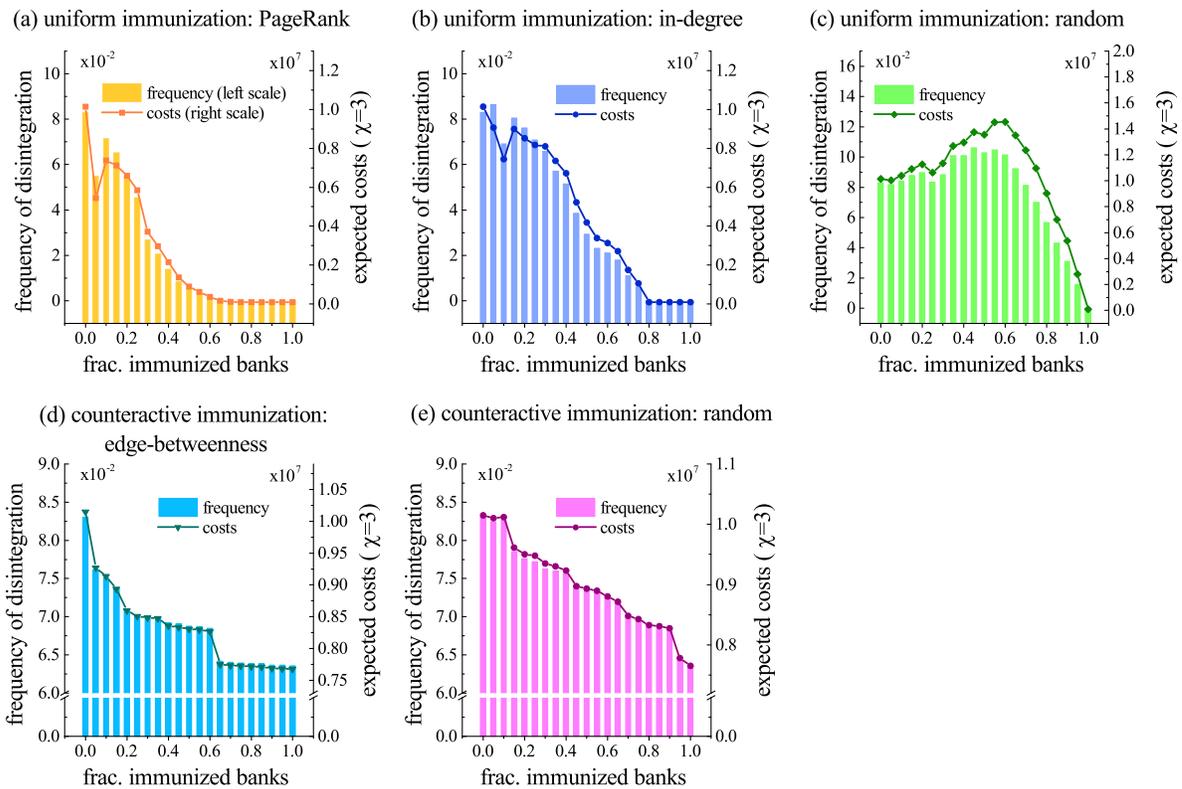}
\caption{{\bf $|$ The frequency with which the GSCC of the financial network disintegrates.} (a)-(c): Uniform immunization. (d) and (e): Counteractive immunization. Restrepo et al.\rq{}s method is used to judge whether the GSCC is present or not.\cite{restrepo08} The frequency of network disintegration is correlated closely with the expected costs under $\chi =3$.  See the caption of Fig.1 for parameter values.}
\end{figure} 

  Fig.\,4 depicts the frequency with which the GSCC of the interbank network disintegrates due to bank defaults (Supplementary Figure S1 shows the simulated percolation threshold). It turns out that uniform immunization tends to decrease the frequency of disintegration, but there is a region in which the likelihood of disintegration is increased. This phenomenon is observed when the de-diversification effect is most profound.

 Under counteractive immunization, on the other hand, the frequency of  disintegration decreases virtually monotonically as vaccination proceeds. This is because counteractive immunization makes it possible to reduce the frequency of medium-scale default cascades without causing the de-diversification effects (Figs.1c,d).

   We also found that the frequency of disintegration follows a trajectory of the expected costs in the case of $\chi \approx 3$. 
 This suggests that the risk of network disintegration could be well captured by the third moment of the number of failed banks. An intuitive explanation for this coincidence is as follows.
Recall that the value of $\chi$ represents the degree of risk aversion, which indicates to what extent simultaneous defaults are more costly than a
sequence of single defaults. If $\chi$= 1, then there is no difference between a simultaneous default of multiple banks and a sequence of single defaults as long as the total number of defaulted banks is the same. 
This implies that the society puts an identical weight on various sizes of collective defaults, which means that the small-scale collective defaults that would not matter for the existence of the GSCC are cared too much.  
When $\chi$ = 10, on the other hand, the society is so risk averse that it cares about infrequent but extremely large-scale simultaneous defaults. The society thus undervalues the costs of the medium-scale collective defaults that would still exceed the percolation threshold.
Only if the value of $\chi$ is just around 3, the society puts appropriate weights on the size of collective defaults  that would be larger than the percolation threshold.

\subsection*{Efficient immunization strategies}

\begin{figure}
\includegraphics[width=15.5cm,clip]{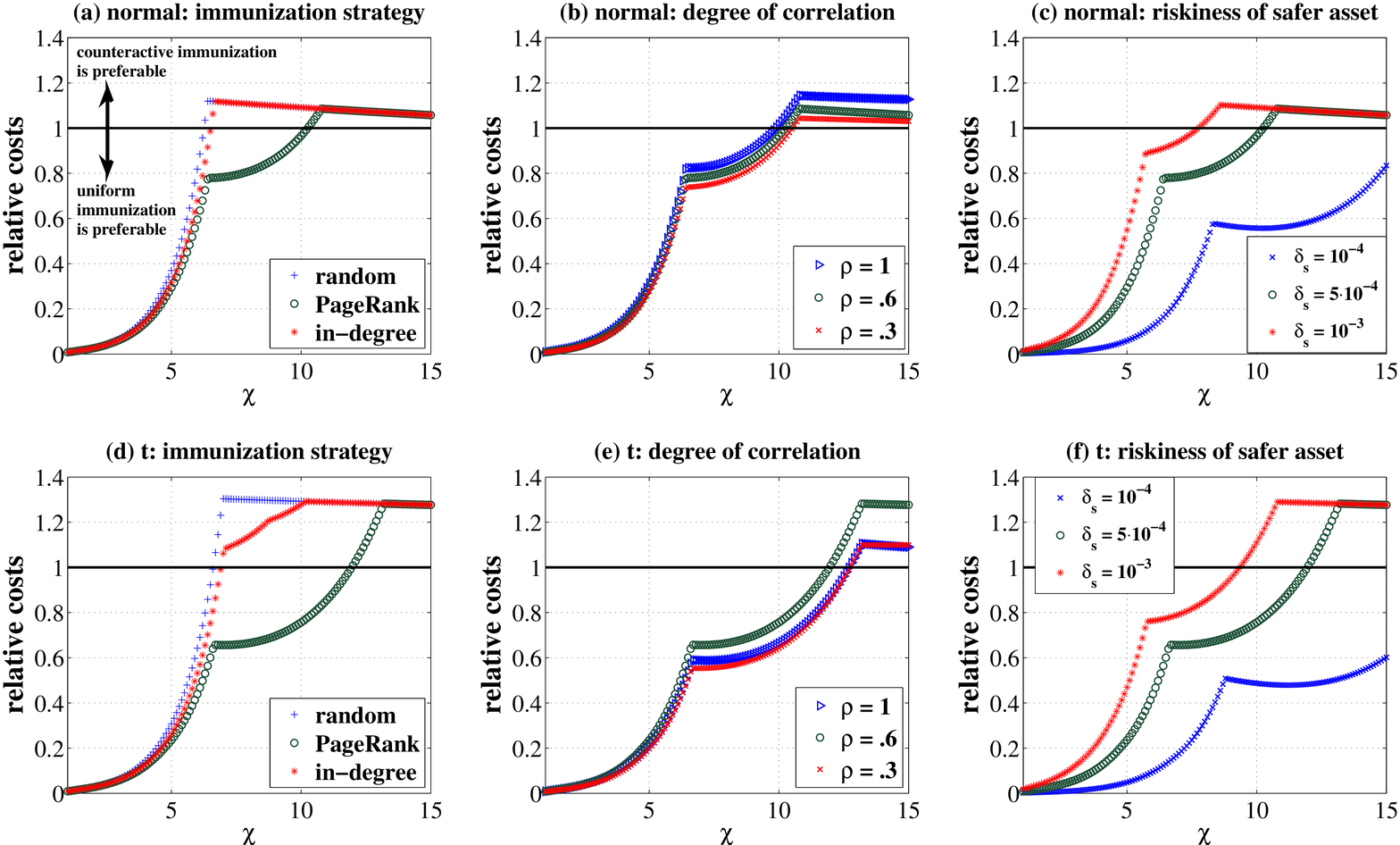}
\caption{{\bf $|$ The expected cost under uniform immunization relative to that under counteractive immunization.} For each immunization strategy, expected costs are minimized with respect to the fraction of immunized banks.
 (a), (b) and (c): asset returns follow a normal distribution. (d), (e) and (f): asset returns follow the t-distribution with degree of freedom 5. Unless otherwise noted, uniform immunization is based on PageRank while counteractive immunization is based on a random order. Baseline parameters are: $\rho = .6$, $\delta_r = 5/N$ and $\delta_s = 1/(2N)$.
 The optimal immunization strategy is to employ uniform immunization when $\chi$ is small  and counteractive immunization when it is large. The threshold value of $\chi$ depends on model parameters.}
\end{figure}

The next question is how to choose an immunization strategy. Fig.5 illustrates the expected costs under uniform immunization relative to that under counteractive immunization. The relative cost depends on various aspects of the model, such as distribution of asset returns, immunization strategy, the degree of asset correlation and the riskiness of the common low-risk asset. 
 
 The main results can be summarized as follows. First, it turns out that uniform immunization is more likely to be preferable as the tail of asset-return distribution becomes thicker. Second, a decrease in $\delta_s$ will move the threshold value of $\chi$ rightward, which means that a counteractive immunization becomes less attractive. Third, the degree of correlation between anti-correlated assets may have an ambiguous impacts on the efficiency of counteractive immunization. Fig.5e illustrates that counteractive immunization is most attractive when $\rho = .6$, while this is not the case if asset returns follow a normal distribution.

A general and robust result is that uniform immunization becomes more efficient than counteractive immunization when $\chi$ is sufficiently small, where the degree of social risk aversion is moderate.
Counteractive immunization becomes more advantageous when $\chi$ is large, where the possibility of large-scale simultaneous defaults is of great concern to the society.

 \section*{Discussion}

 Since systemic risk can be thought of as a fear of default contagion that might occur in financial markets, the previously proposed immunization strategies seem useful for preventing financial contagion. However, a straightforward application of the previous immunization strategies would not work if there is some possibility of serious side effects that would prevail among vaccinated individuals. 

 In the case of uniform immunization, there is a good possibility that a large number of banks will fail simultaneously due to the commonality of assets. We showed that counteractive immunization can be more useful than uniform immunization as long as the society has a strong aversion to such serious side effects. The highest priority for the highly risk-averse society is to avoid  severe financial crises that would rarely occur, rather than a few bank failures that would occur with high probability. Negatively correlated assets can become useful vaccines even if the efficacy of each vaccine is lower than that of the common vaccine that is used under uniform immunization. 

 These results reinforce the idea that micro-prudential policies do not necessarily add up to a macro-prudential policy.\cite{beale11} From a micro point of view, it is efficient to use a common low-risk asset when immunizing banks. From a macro point of view, however, the best immunization strategy may be to prevent a large-scale contagion at the sacrifice of individual failures.  The effectiveness of counteractive immunization suggests that financial networks should be \textit{fragile-yet-robust} rather than \textit{robust-yet-fragile}.

\section*{Methods}

\subsection*{Systemic risk measures}
There is no wide agreement about how to calculate the level of systemic risk. There are many candidates, such as the expected number of defaulted banks, the expected economic losses that would be caused by bank failures, and the likelihood of a large-scale default.\cite{gai10,thurner13,huang13,lenzu12,lorenz09}   
We show two intrinsically different risk measures in order to better understand what is happening in the financial network. The first one is based on the costs that would result from bank failures. 
Following Beale et al.\cite{beale11} and Kobayashi\cite{kobayashi13}, we assume that social costs take the form $C(n) = n^\chi , \chi \in [1, \infty ) $, where $n$ is the number of failed banks. The expected cost is thus given as
$$ E[C] = \sum_{n=1}^N q(n) n^\chi ,$$ 
where $N$ (=1000) is the total number of banks in the financial market and $q(\cdot )$ is the probability function. $\chi$ can be interpreted as the degree of social risk aversion. Notice that if $\chi = 1$, this is simply the average number of defaulted banks. If $\chi > 1$, however, simultaneous defaults of multiple banks become more costly than a sequence of single bank failures. 

 Another measure is the connectivity of the network. As a measure of connectivity, we use the simulated probability that the GSCC of the interbank network disintegrates. In each simulation, we check  whether the GSCC is kept connected after removing the failed banks from the network. We use the method proposed by Restrepo et al.\cite{restrepo08}, who showed that the largest eigenvalue of the adjacency matrix becomes smaller than 1 if the GSCC is absent as long as the network has locally treelike structure. 

  In both of the risk measures, the size of failed banks is not explicitly taken into account. However, this does not necessarily mean that systemic importance is the same across banks. The extent of \lq\lq{}infectivity\rq\rq{} generally varies from bank to bank, depending not only on the amount of borrowings (i.e., in-degree), but also on the topological location in the network and the whole network structure.\cite{kobayashi13} 
Systemic importance of individual banks will differ if their infectivity differs even when their balance-sheet sizes are the same. 
The systemic risk measures used in this study allow us to treat the systemic importance of each bank in an agnostic manner.

\subsection*{Generating an interbank network}
Interbank markets are directed networks. Many studies show that the degree distribution of an interbank network follows a power law.\cite{nier07,soramaki07,iori08,kyriakopooulos09,imakubo10}  
In the present study, the algorithm that generates an interbank network follows from Chung et al.\cite{chung02,chung03}. Their algorithm is also used by Restrepo et al.\cite{restrepo08}.

 Let $d_{in}$ and $d_{out}$ denote the in-degree and out-degree of a bank, respectively. The in-degree of bank $i$ represents the number of banks from which bank $i$ borrows funds. The degree distribution derived by Chung et al.\rq{}s algorithm takes the form $$P(d_l)\propto {d_{l}}^{-\beta},\; l = in, out.$$ We assume that there is no degree-degree correlation. Chung et al.\rq{}s algorithm proceeds as follows. i) Gven the parameters $\beta$, $c$ and $\hat i_0$, express the bank $\hat{i}$\rq{}s in-degree as $d_{in}(\hat i) = c\hat i^{-1/(\beta -1)}$ for $\hat i \in [\hat i_0, \hat i_0\!+\!N]$.
 ii) The out-degree distribution is obtained as a random permutation of the sequence of $d_{in}(\hat i)$. iii) Nodes $\hat i$ and $\hat j (\neq \hat i )$ are linked with probability $d_{out}(\hat i )d_{in}(\hat j)/(dN)$, where $d$ is the average in- or out-degree given by the researcher. iv) By using the condition that the total number of edges is equal to $dN$, it can be shown that $c = [(\beta -2)/(\beta -1)]dN^{1/(\beta -1)}$ and $\hat i_0 = N[(d/m)(\beta -2)/(\beta -1)]^{\beta -1}$ for a sufficiently large $N$, where $m$ denotes the maximum in-degree given by $m = c\hat i_0^{-1/(1-\beta )}$. $m$ and $d$ must be determined in such a way that the probability of placing an edge is less than 1. The parameter values follow from Restrepo et al.\cite{restrepo08} (except for $m$): $\beta = 2.5$, $d=3$ and $m = 50$. To eliminate the possibility that differences in the obtained results might be due to differences in the network structure, we use one particular network structure in all simulations.

\subsection*{The portfolio structure}
Prior to immunization, the asset side of bank $i$'s balance sheet consists of  interbank assets, $l_{i}$, a riskless asset, $b_i$, and a risky external asset, $a_{i}$, where $cov(a_{i},a_{j})=0$ for $j \neq i.$ 
The liability side of bank $i$'s balance sheet consists of interbank liability, $\bar{p}_{i}$, deposits, $d_i$, and
net worth, $w_i$. The balance sheet condition implies that $a_i + b_i + l_i$ = $\bar{p}_i + d_i + w_i$, $i = 1,\ldots , N$. 
The amount of bank $i$'s borrowings from bank $j$ is expressed as $\pi_{ij}\bar{p}_i$, where $\pi_{ij}$ denotes the
relative weight of bank $i$'s borrowings from $j$, and thereby $\sum_{j\neq i} \pi_{ij}=1,\; i = 1,\ldots , N$. 
 The amount of bank $i$'s total interbank assets,
$l_i$, is given by $ l_i  = \sum_{j\neq i}^N \pi_{ji}\bar{p}_j.$ 

 Bank $i$ will default if
 $$
 \bar{p}_i > \sum_{j\neq i}\pi_{ji}{p}_j + \tilde{a}_i + b_i-d_i,
 $$ 
where $\tilde{a}_i$ and $p_j$ stand for the ex-post values of external assets and interbank liabilities, respectively. It should be pointed out that deposits, $d_i$, are reserved because deposits are senior to interbank assets.

Provided that there is no loss in the interbank assets, those banks that hold a common low-risk asset will fail with probability $\delta_s$ while the others will default with probability $\delta_r$.
 Following Beale et al.,\cite{beale11} asset returns are assumed to follow student\rq{}s t-distribution with degree of freedom $v_s$ for the low-risk asset and $v_r$ for the other external assets. For those banks with out-degree 1, the size of net worth, $w_o$, is determined such that $F_{v_r}(-w_o) = \delta_{r}$ or $w_o = -F_{v_r}^{-1}(\delta_r )$, where $F_{v_r}(\cdot )$ is the CDF of student-t distribution with degree of freedom $v_r$. The unit size of the interbank asset, $\bar{l}$, is given as $\bar{l} = \theta_{lw} w_o$, where $\theta_{lw}$ is the predetermined ratio of interbank assets to net worth.  More generally, the amount of net worth is determined as $w_i = l_i/\theta_{lw} = max(1,d_{out}(i)) \bar{l}/\theta_{lw}$ for $i = 1\ldots N$. Given $w_o$, $v_s$ is obtained by solving a nonlinear problem, $F_{v_s}(-w_o) = \delta_s$. In the case where asset returns follow normal distributions (i.e., $v_r \to \infty$), their variances are directly adjusted according to the probability of default, taking as given the size of net worth.

Given the unit size of interbank lending, the sizes of interbank assets and liabilities, $l_i$ and $\bar{p}_i$, are given by the structure of the interbank network. To ensure that the probability of fundamental defaults, i.e., defaults due to the loss of external assets only, is common across banks, the size of external assets relative to the net worth, $\theta_{aw}$, is fixed. 
 If bank $i$ has so many incoming edges that its liability side would be bigger than the sum of interbank assets and external assets, then riskless asset, $b_i$, is imposed to adjust the asset side. Otherwise, deposits, $d_i$, is imposed to adjust the liability side. 

The parameter values are as follows: $v_r = 5$,  $\delta_r = 5/N$, $\theta_{lw} = 3$ and $\theta_{aw}=7$. We use three alternative values of $\delta_s$: $1/N$, $1/(2N)$ and $1/(10N)$. The ratio of interbank assets to capital is roughly consistent with the data shown by Upper.\cite{upper11} These parameters indicate  that the ratio of capital to total risky assets is .1 for those banks that have a positive amount of interbank assets and $1/7 \approx .143$ for the other banks.
Solving the above-mentioned nonlinear problem yields $v_s = 10.930$ for $\delta_s = 1/N$, 15.687 for $\delta_s = 1/(2N)$, and 47.296 for $\delta_s = 1/(10N)$.
 Random asset returns are generated $10^5$ times for each case.

\subsection*{The algorithm for detecting failed banks} If bank $i$ failed, then bank $i$\rq{}s creditors lose $k\%$ of their credits extended to bank $i$. Some of these creditors may fail due to the loss of their interbank assets. Accordingly, the creditors of the creditors of bank $i$ may fail as well, because they lose $k\%$ of their credits extended to the failed banks. The spread of contagious defaults stops if it turns out that no more banks are going to fail due to the loss of interbank assets. Following the literature, we assume that $k = 100$.

\subsection*{Immunization strategies}
  Under uniform immunization, the vaccination order is based on either PageRank\cite{brin98} or in-degree.  
PageRank, $y_i$, is given by
$  y_i = \alpha \sum_j A_{ij}{y_j}/{k_j^{out}} + \beta ,$
where $A_{ij}$ is the $(i,j)$-th element of the adjacency matrix, which takes 1 if $\pi_{ij}>0$ and 0 otherwise. $k_j^{out}$ is the out-degree of node $j$. The parameter values are $\alpha =.85$ and $\beta = 1$. When two or more nodes have a tie score, we randomize the order among them.
 
 Under counteractive immunization, we pick up a pair of banks sequentially. In the case of the edge-betweenness-based strategy, we first select a pair of banks whose edge spanned between the two banks takes the highest value in terms of the edge-betweenness measure.\cite{rubinov10}  We remove the edges coming from and going to the selected nodes, as well as the selected nodes themselves. In the following step, edge-betweenness centrality is recalculated, and then the new highest-ranked edge is selected. If there are multiple edges that have the same edge-betweenness, we randomly choose one edge among those edges. 
As for random strategy, we basically randomize the vaccination order obtained by the edge-betweenness-based strategy. The only difference is that we randomly choose a new pair of banks when there are multiple edges that exhibit the same edge-betweenness. In both strategies, we randomly pick up a pair of banks from the remaining banks if there are no more edges in the network.
  
The selected banks are required to hold negatively correlated external assets whose correlation coefficient is $-\rho$. Any two correlated returns, $\{\varepsilon_i, \varepsilon_j\}, i\neq j $, are expressed as
$$ \{ \varepsilon_i ,  \varepsilon_j \} = \{ \varepsilon_i ,  -\rho \varepsilon_i + (1-\rho )\hat{\varepsilon}_j    \}, i \neq j,
$$
where $\varepsilon_s$ has mean zero and variance $\sigma^2$, while $\hat{\varepsilon}_s$ has mean zero and variance $(1+\rho )\sigma^2 /(1-\rho )$. There is no correlation between $\varepsilon_i$ and $\hat{\varepsilon}_j$, $\forall\: j \neq i $.
Notice that the variance of each asset return is independent of the value of $\rho$, which means that the riskiness of individual assets remains unchanged.

\section*{Acknowledgments} 
Kobayashi gratefully acknowledges financial support from KAKENHI 25780203 and 24243044. 

\section*{Author contribution}
T.K. designed research, performed research and wrote the paper. T.K. and K.H. wrote the Matlab codes, and K.H. surveyed the literature and prepared the figures.

\end{document}